\documentclass[journal,twoside,web]{ieeecolor}
\usepackage{generic}
\usepackage{cite}
\usepackage{amsmath,amsfonts,amssymb}
\usepackage{algorithm}
\usepackage{algorithmicx}
\usepackage{algpseudocode}
\usepackage{array}
\usepackage{textcomp}
\usepackage{url}
\usepackage{graphicx}
\usepackage{booktabs}
\usepackage{tabularx}

\usepackage{multirow}
\usepackage{makecell}
\usepackage{bm}
\usepackage[dvipsnames]{xcolor}
\usepackage[colorlinks=true,linkcolor=Cerulean,citecolor=Cerulean,urlcolor=Cerulean]{hyperref}
\usepackage{orcidlink}
\usepackage{cleveref}
\crefformat{figure}{Fig.~#2#1#3}
\crefformat{table}{Table~#2#1#3}
\crefformat{algorithm}{Algorithm~#2#1#3}
\crefformat{section}{#2Section~#1#3}
\crefformat{subsection}{#2Section~#1#3}
\usepackage{placeins}

\begin{document}

\title{LLM-Guided Measurement Credibility Correction for Trustworthy Industrial Process Inference}

\author{
  Youcheng Zong\(^{\orcidlink{0009-0008-6795-8412}}\),~\IEEEmembership{Student~Member,~IEEE},
  Runda Jia\(^{\orcidlink{0000-0002-8586-243X}}\),
  \\ and Dakuo He\(^{\orcidlink{0000-0001-8303-529X}}\)
  \thanks{This work was supported by the Fundamental Research Funds for the Central Universities, China (N26GFZ006). \textit{(Corresponding author: Runda Jia.)}}
  \thanks{Youcheng Zong, Runda Jia, and Dakuo He are with the College of Information Science and Engineering, Northeastern University, Shenyang 110004, China (e-mail: youchengzong@stumail.neu.edu.cn; jiarunda@ise.neu.edu.cn; hedakuo@ise.neu.edu.cn).}
}

\maketitle

\begin{abstract}
  Industrial prediction and soft sensing depend on credible input measurements. In field deployment, a predictor may receive biased, delayed, stale, or derived measurements that still look plausible. Prediction can then fail before the forecasting backbone becomes the main limitation, because the input window no longer represents the real process.
  Sensor reconstruction, data reconciliation, and fault-tolerant soft sensing reduce this risk, but they often rely on numerical correlation, alarms, fault labels, or explicit process equations. These assumptions are not always available. A correlated variable can also be an unsafe reference when variables share instruments, derived formulas, soft-sensing chains, or control actions. The key issue is to decide before prediction which external measurements can credibly support the current measurement.
  To address this issue, this article proposes LLM-Guided Measurement Credibility Correction (MCC). MCC converts measurement meanings in process documents into measurement semantics usable by numerical models. It builds independent process references from semantically qualified external measurements and corrects local measurement conflicts before prediction. The predictor therefore receives a more credible input window.
  Across multiple complex industrial forecasting and soft-sensing tasks, +MCC achieves average relative MAE reductions of 30.7\% on real-test protocols and 80.3\% on controlled-corruption protocols. It adds only 0.5--2.0k online parameters, with the slowest +MCC inference time at 0.089 ms/step. These results show that measurement semantics can turn process documents into lightweight pre-inference credibility correction and improve prediction accuracy.
\end{abstract}

\begin{IEEEkeywords}
  large language models, measurement credibility correction, industrial process inference, soft sensing, measurement semantics.
\end{IEEEkeywords}

\section{Introduction}\label{sec:introduction}

Process industries rely on online prediction and soft sensing for monitoring, scheduling, optimization control, and operator decisions. These models usually receive multivariate measurement windows and infer product indicators, process indicators, or equipment states that cannot be measured fast enough or directly. Studies on soft sensing and transfer soft sensing have improved how data-driven models learn industrial dynamics and transfer across operating conditions~\cite{sun2021survey,liu2026transfersoftsensor}. In real deployment, a predictor does not receive abstract numerical columns. It receives on-site measurements with measurement sources, units, sampling modes, and process roles. Thus, the reliability of industrial prediction also depends on the credibility of the input measurements themselves.

This need is most visible when a measurement is distorted without triggering an alarm. The measured value may look normal but no longer represent the real process. Flow-meter drift, temperature lag, upstream instrument errors, derived calculations, and sampling delays can locally distort the input window. Sensor reconstruction and fault-tolerant soft sensing have used external variables to recover abnormal channels or improve model tolerance to abnormal inputs~\cite{yue2001reconstruction,zhang2025faulttolerantsoftsensor}. The predictor, however, still needs a credible input window. Thus, prediction should start from credible inputs. The model must identify unreliable measurements before inference and correct distorted inputs only when reliable evidence is available.

The key question is what evidence can be trusted. A correlated variable is not always a credible reference. Two variables may be correlated because of a real process relation, but they may also be jointly distorted by a shared instrument, a derived formula, a soft-sensing chain, or a control action. Separating these two cases requires knowing how each variable is measured and how it is used in the process. Variable tables and process documents record variable names, units, measurement modes, sampling periods, and process descriptions, but this information is usually open-text and semi-structured. Ordinary numerical models cannot directly use it. LLMs can use world knowledge and logical reasoning to extract auditable measurement semantics from these texts. These semantics allow the model to decide which variables can provide credible references for another variable. Studies on LLM-based time series show that language information can provide useful priors for time-series modeling and support more flexible numerical-sequence modeling interfaces~\cite{jin2024timellm,liu2024autotimes}. This ability moves measurement credibility judgment from purely numerical correlation toward semantically qualified external references.

\begin{figure}[t]
  \centering
  \includegraphics[width=88mm]{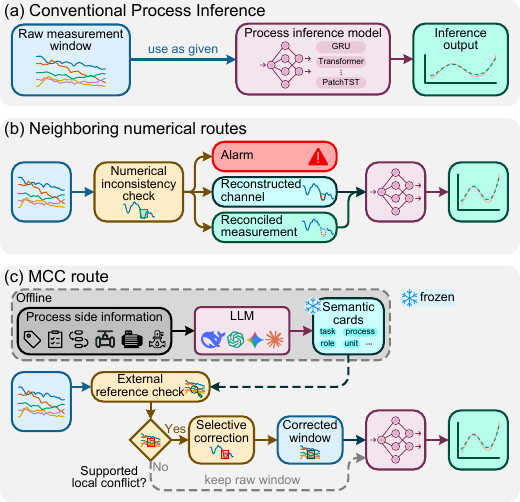}
  \caption{MCC problem setting and neighboring routes. (a) Raw measurement windows are fed directly to the predictor. (b) Neighboring numerical routes handle inconsistency through alarms, reconstruction, or reconciliation. (c) MCC builds measurement semantics offline and uses LLM-free external references online to decide whether to correct local conflicts.}
  \label{fig:intro_mcc}
\end{figure}

This article proposes LLM-Guided Measurement Credibility Correction (MCC). Before training, MCC uses an LLM's world knowledge and logical reasoning to turn variable documents and process text into frozen measurement semantics. These semantics guide the numerical observer in selecting external references.
In the online stage, the observer selects qualified external evidence for each correctable variable and builds a process reference that does not use the variable itself. If the current measurement has a local conflict with a reliable reference, MCC corrects it before prediction. If the reference is unreliable, or if many variables jointly reflect a real operating change, MCC remains conservative. \cref{fig:intro_mcc} illustrates this setting.

The main contributions are as follows.
\begin{itemize}
  \item
        We propose MCC, which uses LLM-derived measurement semantics to judge before prediction whether input measurements still credibly represent the current process.
  \item
        We introduce measurement-semantic references, which allow the model to find independent evidence from qualified external measurements and conservatively correct supported local input conflicts before prediction.
  \item
        On multiple complex industrial forecasting and soft-sensing tasks, MCC improves performance across different backbones, with negligible additional online inference overhead.
\end{itemize}
The remainder of this paper is organized as follows. \cref{sec:related_work} reviews related work. \cref{sec:method} presents the method. \cref{sec:experiments} reports the experiments. \cref{sec:conclusion} concludes the paper.

\section{Related Work}\label{sec:related_work}

\subsection{Measurement Credibility in Industrial Prediction}

Industrial prediction usually starts from historical measurement windows. Recurrent networks and Transformers have been used to model nonlinear temporal relations in industrial dynamics~\cite{yuan2020lstm,geng2022transformer}, and data-driven models have also been applied to industrial tasks such as material-pressure prediction~\cite{zong2025hybridgrid}. Process-graph-constrained soft sensing further introduces structural priors into the predictor, so that the model does not rely only on anonymous numerical channels~\cite{zhai2024processgraphsoftsensor}. These studies show that suitable models can learn temporal dynamics and process structures from multivariate windows.

These advances mainly improve how the model uses the measurement window, but they rarely check whether the window still represents the current process. In field deployment, the measurement chain may first distort part of the input, and the model then predicts from that input. A stronger temporal model may still infer from wrong evidence. MCC therefore places measurement credibility before prediction, so that the input window is checked before it enters the forecasting model.

\subsection{Measurement Validation and Reconstruction}

Inconsistent measurements have long been central to measurement validation, process monitoring, and fault diagnosis. Process fault detection and diagnosis~\cite{venkatasubramanian2003fddquantitative}, data-driven process monitoring~\cite{qin2012processmonitoring}, sensor reconstruction~\cite{yue2001reconstruction}, and data reconciliation~\cite{narasimhan1999datareconciliation} address this problem through fault detection and diagnosis, process history, external variables, and constraint consistency, respectively. Sensor reconstruction recovers abnormal readings from other variables, while data reconciliation and gross error detection usually use process equations or balance constraints to correct inconsistent measurements.

These routes show that measurement correction needs independent evidence and cannot rely only on the channel being corrected. Pre-inference correction also requires an output that can serve downstream prediction, rather than only an alarm, a fault type, or an offline diagnosis conclusion. When explicit process equations, fault labels, or alarm information are unavailable, usable evidence must come from other online measurements. MCC organizes these measurements into credible external references and uses them for input correction before prediction.

\subsection{Variable Relations and Measurement Semantics}

Deep models for multivariate monitoring can learn statistical relations among sensors. Graph neural networks explicitly model variable dependencies and have been used for multivariate anomaly detection~\cite{deng2021gnnad,zheng2024adversarialgnn}. Such relations are useful for monitoring and reconstruction, but correlation itself does not show that one variable can serve as a credible reference for another. In an industrial measurement chain, correlation may come from real process coupling, but it may also come from a shared instrument, a derived formula, a control action, or a sampling delay. Reference construction therefore needs to judge whether a variable is qualified to provide independent support.

Deciding whether a variable is qualified requires information beyond the numerical window, including variable names, units, measurement sources, sampling modes, derived relations, and process roles. LLM-based and text-enhanced time-series studies provide interfaces for using language information, including pre-trained model reprogramming, cross-modal matching, and text prototypes~\cite{jin2024timellm,liu2025timecma,sun2024test}. Industrial studies also use LLMs or knowledge structures for causal representation, human-AI decision support, zero-shot diagnosis, and knowledge-graph-assisted diagnosis~\cite{yao2026causalllm,zong2026llmdecision,han2026zeroshot,liu2024kgllm,su2025zero}. These studies show that textual or semantic knowledge can supply context missing from purely numerical models. However, directly placing an LLM inside the forecasting loop adds cost and does not always bring stable gains~\cite{merrill2024llmts}. MCC uses a lighter interface: measurement semantics are constructed before training, while online reference construction and input correction remain numerical.

\section{Method}\label{sec:method}

\subsection{Problem Formulation}\label{sec:method_problem}
Let the multivariate industrial series be $X_{1:T}=\{\mathbf{x}_t\}_{t=1}^{T}$, where $\mathbf{x}_t=[x_1(t),\ldots,x_N(t)]^\top$. At time $t$, the predictor receives a measurement window $W_t=X_{t-L+1:t}$ of length $L$ and predicts the next target $y_{t+1}$.
\begin{equation}
  \mathbf{x}_t=
  [x_1(t),\ldots,x_N(t)]^\top,
  \qquad
  W_t=X_{t-L+1:t}.
\end{equation}
Continuous measurement variables are standardized with the training-set mean $\mu_i$ and standard deviation $\sigma_i$. For the $i$-th variable, the standardized measurement is
\begin{equation}
  \bar{x}_i(t)=\frac{x_i(t)-\mu_i}{\sigma_i}.
\end{equation}
The window formed by these standardized measurements is denoted by $\bar{W}_t$. Here, $x_i(t)$ is not assumed to be a reliable observation of the process state; it is measurement evidence that should be checked before prediction. MCC first corrects $\bar{W}_t$ into $\bar{W}_t^c$ and then sends it to the downstream predictor.

Not every input column is a proper correction target. Let $\mathcal{M}\subseteq\{1,\ldots,N\}$ denote the set of correctable measurement variables, which is fixed before training from variable types and metadata. Time indices, stage indicators, and other pure context variables may provide evidence for other variables, but they are not correction targets.
For each $i\in\mathcal{M}$, MCC checks whether the current measurement $\bar{x}_i(t)$ is still credible. When the external evidence is reliable enough, MCC corrects this measurement toward an independent process reference $\bar{x}_i^*(t)$ before prediction.

\subsection{Offline Measurement Semantics}\label{sec:method_semantics}
Before training, MCC calls the LLM to convert the industrial scenario and variable information into measurement semantics. The side information is denoted by $D$. It consists of available scenario descriptions, variable names, units, sampling periods, measurement sources, and process descriptions. If process documents are unavailable, MCC uses the existing variable information. Based on $D$, the LLM analyzes each variable $x_i$ and generates its measurement-semantic description $z_i$:
\begin{equation}
  z_i=\Phi_{\mathrm{LLM}}(D,i),
  \qquad i=1,\ldots,N.
\end{equation}
The description $z_i$ records the measurement source, process role, control or observation attribute, sampling relation, upstream or downstream position, and derived relation. It also records redundant variables that can provide independent references. MCC uses these semantics to judge which external measurements are qualified to support the target variable.

\begin{figure*}[t]
  \centering
  \includegraphics[width=181.8mm]{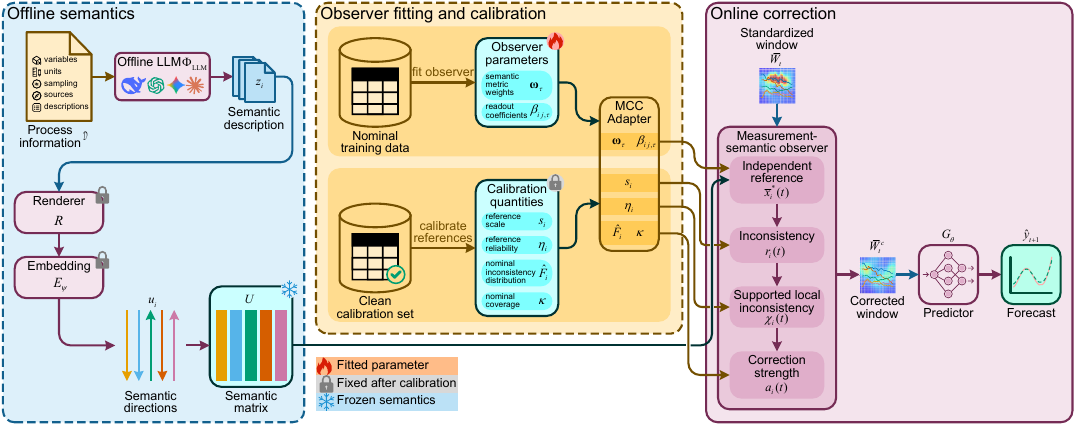}
  \caption{MCC method overview. Left: offline LLM processing builds the frozen measurement semantics $U$ from $D$. Middle: nominal training fits the observer, and clean calibration fixes $s_i$, $\eta_i$, and $\widehat{F}_i$. Right: online correction computes independent references and correction strengths without LLM calls before sending $\bar{W}_t^c$ to $G_\theta$.}
  \label{fig:mcc_method_overview}
\end{figure*}

A fixed text renderer $R$ and a fixed embedding model $E$ convert $z_i$ into a measurement-semantic direction $u_i$:
\begin{equation}
  u_i=\operatorname{norm}\!\left(E(R(z_i))\right)\in\mathbb{R}^{k}.
\end{equation}
Stacking all directions in input-column order gives the frozen semantic matrix $U$:
\begin{equation}
  U=
  \begin{bmatrix}
    u_1^\top\\
    \cdots\\
    u_N^\top
  \end{bmatrix}
  \in\mathbb{R}^{N\times k}.
\end{equation}
$U$ remains fixed during training, calibration, and inference. The LLM does not receive the online window $W_t$, and it does not output credibility scores, fault labels, rules, thresholds, or correction values.
\cref{fig:mcc_method_overview} gives an overview of the MCC method.

\subsection{Measurement-Semantic Reference}\label{sec:method_reference}
For the target variable $x_i$, the independent process reference is built only from other variables and does not access $x_i(t)$ itself. Let $H\le L$ denote the number of lags used by the observer. For a candidate variable $x_j$ and lag $\tau$, the semantic activation is
\begin{equation}
  e_{j,\tau}(t)=\bar{x}_j(t-\tau)u_j,
  \qquad
  j\ne i,\quad \tau=0,\ldots,H-1.
\end{equation}
Here, $\bar{x}_j(t-\tau)$ gives the current numerical strength, and $u_j$ gives the measurement-semantic direction of that variable.

MCC learns a semantic weight matrix $\Omega_\tau$ for each lag, where $\boldsymbol{\omega}_\tau$ is learned from nominal training data and shared by all variables at the same lag:
\begin{equation}
  \Omega_\tau=\operatorname{diag}(\boldsymbol{\omega}_\tau).
\end{equation}
The semantic compatibility between the target variable $x_i$ and the candidate variable $x_j$ at lag $\tau$ is
\begin{equation}
  c_{ij,\tau}=u_i^\top\Omega_\tau u_j,
  \qquad j\ne i.
\end{equation}
The routing weights are determined by the target semantic direction, the candidate semantic directions, and the lags:
\begin{equation}
  (\alpha_{ij,\tau})_{j\ne i,\,\tau}
  =
  \rho_i\!\left((|c_{ij,\tau}|)_{j\ne i,\,\tau}\right),
  \qquad
  \alpha_{ij,\tau}\ge0,\quad
  \sum_{\tau=0}^{H-1}\sum_{j\ne i}\alpha_{ij,\tau}=1.
\end{equation}
Here, $\rho_i$ is a sparse normalization map over candidate variable--lag pairs. It maps nonnegative compatibilities to nonnegative weights and enforces the unit-sum constraint. When top-$k$ support sparsity is used, the sparsity level is applied by $\rho_i$ before normalization. This routing does not use the current value of $x_i(t)$, so the checked measurement cannot support itself.

The external semantic support strength is
\begin{equation}
  g_i=
  \sum_{\tau=0}^{H-1}
  \sum_{j\ne i}
  \alpha_{ij,\tau}|c_{ij,\tau}|.
\end{equation}
The independent reference of variable $x_i$ is
\begin{equation}
  \bar{x}_i^*(t)
  =
  \sum_{\tau=0}^{H-1}
  \sum_{j\ne i}
  \alpha_{ij,\tau}\,
  \beta_{ij,\tau}\,
  c_{ij,\tau}\,
  \bar{x}_j(t-\tau).
  \label{eq:mcc_reference}
\end{equation}
The coefficient $\beta_{ij,\tau}$ is an unbiased numerical readout coefficient. Semantic compatibility and routing decide which external measurements are qualified to support $x_i$; $\beta_{ij,\tau}$ only aligns these qualified sources to the standardized scale of $x_i$ and does not introduce a free bias.
Equivalently, the observer first aggregates semantic activations into $m_i(t)\in\mathbb{R}^k$ and then reads out along the target direction:
\begin{equation}
  m_i(t)=
  \sum_{\tau=0}^{H-1}
  \sum_{j\ne i}
  \alpha_{ij,\tau}\,
  \beta_{ij,\tau}
  \Omega_\tau e_{j,\tau}(t),
  \qquad
  \bar{x}_i^*(t)=u_i^\top m_i(t).
\end{equation}
This form preserves leave-one-out independence: the target semantics $u_i$ query and read out external evidence, but the current target measurement itself does not enter the reference construction.

On the clean calibration set, MCC estimates the nominal residual scale $s_i$ of each reference by the median of $|\bar{x}_i(t)-\bar{x}_i^*(t)|$ and obtains the reference reliability
\begin{equation}
  \eta_i=\frac{1}{1+s_i}.
\end{equation}
The support weight considers both semantic support and numerical reliability:
\begin{equation}
  p_i=g_i\eta_i.
\end{equation}
A smaller $s_i$ gives a larger $\eta_i$, indicating that the reference is more stable on nominal data. A smaller $p_i$ means that the current variable lacks a stable external reference, so the later correction is naturally weakened. The same $\eta_i$ also limits the correction magnitude in the final update.

\subsection{Supported Inconsistency and Correction}\label{sec:method_correction}
The standardized inconsistency between the current measurement and its independent reference is
\begin{equation}
  r_i(t)=
  \frac{|\bar{x}_i(t)-\bar{x}_i^*(t)|}{s_i},
  \qquad i\in\mathcal{M}.
\end{equation}
The support-weighted inconsistency is
\begin{equation}
  d_i(t)=p_i r_i(t).
\end{equation}
Raw inconsistency alone is not enough to trigger correction. A real operating change may make many variables deviate from their references at the same time. MCC therefore subtracts the background inconsistency from the target inconsistency:
\begin{equation}
  \chi_i(t)=
  d_i(t)
  -
  \operatorname{median}_{\ell\in\mathcal{M},\,\ell\ne i}d_\ell(t).
  \label{eq:mcc_supported_inconsistency}
\end{equation}
$\chi_i(t)$ denotes supported local inconsistency. It increases when one variable conflicts with a reliable external reference, and it is suppressed by the background term when many variables change together.

Let $\mathcal{C}$ be the set of clean calibration time indices. The empirical distribution of nominal inconsistency for variable $x_i$ is
\begin{equation}
  \widehat{F}_i(s)=
  \frac{1}{|\mathcal{C}|}
  \sum_{n\in\mathcal{C}}
  \mathbf{1}\{\chi_i(n)\le s\}.
\end{equation}
The position of the current inconsistency in this nominal distribution is
\begin{equation}
  q_i(t)=\widehat{F}_i(\chi_i(t)).
\end{equation}
Let $\kappa\in(0,1)$ denote nominal coverage. It sets the intended fraction of clean calibration states that should not trigger correction. The default experiment setting is $\kappa=0.95$. The correction strength is
\begin{equation}
  a_i(t)=
  \left[
  \frac{q_i(t)-\kappa}{1-\kappa}
  \right]_0^1,
  \qquad
  [z]_0^1=\min(\max(z,0),1).
  \label{eq:mcc_correction_strength}
\end{equation}
If $q_i(t)\le\kappa$, then $a_i(t)=0$. When the local conflict is sufficiently rare under the calibration distribution, $a_i(t)$ increases with its nominal rarity.

\begin{figure}[t]
  \centering
  \includegraphics[width=88mm]{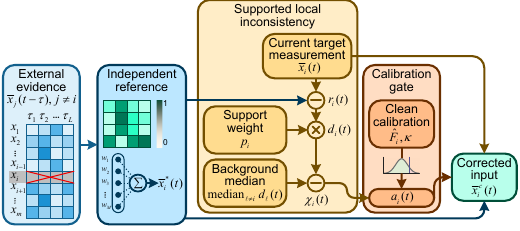}
  \caption{Measurement-semantic reference and local correction path. The target measurement $\bar{x}_i(t)$ is excluded from the evidence set, so $\bar{x}_i^*(t)$ is built only from external variables and their lags. The right side shows how support weighting, background subtraction, and calibration convert $r_i(t)$ into a mixing weight.}
  \label{fig:mcc_reference_correction}
\end{figure}

For $i\in\mathcal{M}$, the corrected standardized input is
\begin{equation}
  \bar{x}_i^c(t)=
  (1-\eta_i a_i(t))\bar{x}_i(t)
  +
  \eta_i a_i(t)\bar{x}_i^*(t).
  \label{eq:mcc_corrected_input}
\end{equation}
For $i\notin\mathcal{M}$, MCC keeps $\bar{x}_i^c(t)=\bar{x}_i(t)$. These variables may enter the observer as evidence, but they are not continuously corrected.
Thus, when the measurement lies in the nominal range, $a_i(t)\approx0$ and $\bar{x}_i^c(t)\approx\bar{x}_i(t)$. When a local distortion is supported by a reliable external reference, $\bar{x}_i^c(t)$ moves toward $\bar{x}_i^*(t)$. Variables without such references are not forced to be corrected. \cref{fig:mcc_reference_correction} gives the corresponding local update path.

\subsection{Training and Inference}\label{sec:method_training}
Training has three ordered steps. MCC first learns the measurement-semantic observer, then calibrates the inconsistency distribution, and finally trains the downstream predictor with the observer and calibrated quantities fixed. The semantic matrix $U$ remains frozen during all three steps. The observer parameters are $\Theta_{\mathrm{obs}}=\{\boldsymbol{\omega}_\tau,\beta_{ij,\tau}\}$.
Let $\mathcal{T}_{\mathrm{obs}}$ denote the nominal observer-training index set. The observer is learned by fitting the independent references of correctable measurement variables:
\begin{equation}
  \Theta_{\mathrm{obs}}^\ast
  =
  \arg\min_{\Theta_{\mathrm{obs}}}
  \sum_{t\in\mathcal{T}_{\mathrm{obs}}}
  \sum_{i\in\mathcal{M}}
  \left(\bar{x}_i^*(t)-\bar{x}_i(t)\right)^2 .
  \label{eq:mcc_observer_training}
\end{equation}
This objective uses only nominal data and does not require fault labels. After fitting, MCC estimates $s_i$, $\eta_i$, and $\widehat{F}_i$ on the clean calibration set $\mathcal{C}$, and keeps them fixed during the later predictor training and inference stages.

The downstream predictor is denoted by $G_\theta$. It receives the corrected window and outputs the prediction:
\begin{equation}
  \hat{y}_{t+1}=G_\theta(\bar{W}_t^c).
  \label{eq:mcc_prediction}
\end{equation}
Let $\mathcal{T}_{\mathrm{pred}}$ denote the predictor-training index set. During training, the fitted and calibrated MCC module generates $\bar{W}_t^c$ from $\bar{W}_t$. Given the supervised loss $\mathcal{L}_{\mathrm{pred}}$, the predictor parameters are learned by empirical risk minimization:
\begin{equation}
  \theta^\ast=
  \arg\min_{\theta}
  \frac{1}{|\mathcal{T}_{\mathrm{pred}}|}
  \sum_{t\in\mathcal{T}_{\mathrm{pred}}}
  \mathcal{L}_{\mathrm{pred}}
  \left(G_\theta(\bar{W}_t^c),y_{t+1}\right).
  \label{eq:mcc_predictor_training}
\end{equation}
The predictor can be any time-series model that accepts a multivariate window. MCC and the baseline use the same data split, normalization statistics, prediction model, and supervised loss. The only difference is that the baseline receives $\bar{W}_t$, whereas MCC receives $\bar{W}_t^c$.
At inference, the LLM is not called. MCC computes the semantic references, inconsistency, correction strength, and corrected window in order. The predictor $G_\theta$ then predicts $y_{t+1}$. The test set is used only for final evaluation. It is not used to construct semantics, estimate normalization statistics, fit the observer, or select the calibration distribution.

\section{Experiments}\label{sec:experiments}

This section evaluates MCC on three industrial process-inference tasks.

\subsection{Experimental Settings}\label{sec:4-1}

All methods are evaluated in the same implementation and under the same training protocol. For each dataset and backbone, Base and +MCC share the same split, normalization statistics, backbone configuration, training budget, and model-selection rule. The test set is not used for tuning, observer fitting, or calibration.
Measurement semantics are constructed offline before training and then fixed for training and inference. The construction protocol uses available industrial-scenario information, variable names, and metadata to form the measurement meaning and process role of each variable.
All experiments are run on a workstation with an Intel Xeon 8470Q CPU, 90\,GB RAM, and a single NVIDIA RTX 5090 GPU (32\,GB). The implementation uses Python~3.14, PyTorch~2.11.0, and CUDA~13.0. Quantitative results are averaged over three independent runs.
The code repository is available at \href{https://github.com/mituan-ai/MCC_open}{https://github.com/mituan-ai/MCC\_open}.
\cref{tab:experiment_config} summarizes the common experimental settings.

\begin{table}[t]
  \centering
  \caption{Common training, evaluation, and MCC settings. The table separates the shared backbone protocol from MCC-specific semantics, observer, and correction-gate settings.}
  \label{tab:experiment_config}
  \setlength{\tabcolsep}{10pt}\tiny
  \begin{tabular}{lll}
    \toprule
    Name                    & \multicolumn{2}{l}{Setting}                                    \\
    \midrule
    Optimizer               & \multicolumn{2}{l}{AdamW}                                      \\
    Learning rate           & \multicolumn{2}{l}{10\textsuperscript{-3}}                     \\
    Loss                    & \multicolumn{2}{l}{MSE}                                        \\
    Max epochs              & \multicolumn{2}{l}{300}                                        \\
    Early stopping          & patience                                  & 40                \\
    \multirow[t]{3}{*}{LR scheduler}
                            & type                                       & ReduceLROnPlateau \\
                            & decay factor                               & 0.5               \\
                            & patience                                   & 15                \\
    Weight decay            & \multicolumn{2}{l}{10\textsuperscript{-4}}                     \\
    Gradient clipping       & \multicolumn{2}{l}{1.0}                                        \\
    Batch size              & \multicolumn{2}{l}{64}                                         \\
    Measurement semantics   & LLM                                        & DeepSeek-v4-pro   \\
                            & reasoning effort                           & max               \\
    Embedding model         & \multicolumn{2}{l}{text-embedding-v4}                          \\
    Semantic dimension      & $k$                                        & 128               \\
    Observer lags           & $H$                                        & 3                 \\
    Support sparsity        & top-$k$                                    & 6                 \\
    Observer fitting        & epochs                                    & 3                 \\
                            & learning rate                             & 5$\times$10\textsuperscript{-3} \\
    Correction coverage     & $\kappa$                                   & 0.95              \\
    \bottomrule
  \end{tabular}
\end{table}

Prediction performance is evaluated by mean absolute error (MAE), root mean squared error (RMSE), and coefficient of determination ($R^2$). Online cost is evaluated by per-step inference time, parameter count, and peak RSS memory.
The comparison covers eight common time-series backbones: GRU, LSTM, Transformer, Informer, Mamba, iTransformer, PatchTST, and ModernTCN. Among them, Transformer, Mamba, iTransformer, PatchTST, and ModernTCN represent classical and recent sequence-modeling routes~\cite{vaswani2017transformer,gu2024mamba,liu2024itransformer,nie2023patchtst,luo2024moderntcn}. Base and +MCC receive the same measurement window; +MCC additionally uses the fixed pre-training measurement semantics to construct the observer and correct the input window before prediction.

\subsection{Datasets}\label{sec:4-2}

The experiments use three industrial process datasets, covering ladle-preheating temperature forecasting, thickener underflow-concentration estimation, and fed-batch penicillin-concentration estimation. Each dataset contains two test protocols. Real Test is the original fixed test split without additional artificial measurement corruption. Corrupted Test injects controlled measurement corruption into specified real-test subsets to evaluate pre-inference measurement correction.
All training, validation, and test splits are fixed before training at the process, scenario, or batch level. \cref{tab:dataset_protocol} summarizes the inputs, targets, splits, and Corrupted Test scale for each dataset. Real Test covers industrial deployment cases such as complete-process holdout, far-domain regimes, and fault batches.

\begin{table}[t]
  \centering
  \caption{Dataset protocols for the MCC experiments. Each block gives the input window, prediction target, fixed splits, and Corrupted Test size for one process.}
  \label{tab:dataset_protocol}
  \setlength{\tabcolsep}{2pt}\tiny
  \begin{tabular}{p{0.1\linewidth}p{0.6\linewidth}rl}
    \toprule
    Item   & \multicolumn{3}{l}{Description}                                                       \\
    \midrule
    \multicolumn{4}{l}{\textit{Ladle Preheating}}                                                  \\
    \multirow[t]{3}{*}{Input}
           & Variables                                                          & 14     &         \\
           & Sampling interval                                                  & 5      & minutes \\
           & Input window                                                       & 60     & steps   \\
    Target & \multicolumn{3}{l}{Ladle temperature over the next 5 steps}                           \\
    \multirow[t]{4}{*}{Splits}
           & 24 training processes                                              & 15,236 & samples \\
           & 5 validation processes                                             & 2,500  & samples \\
           & 3 held-out test processes                                          & 2,104  & samples \\
           & 14 corrupted test cases                                            & 29,456 & samples \\
    \midrule
    \multicolumn{4}{l}{\textit{Thickener Dewatering}}                                              \\
    \multirow[t]{3}{*}{Input}
           & Variables                                                          & 5      &         \\
           & Sampling interval                                                  & 1      & minute  \\
           & Input window                                                       & 30     & steps   \\
    Target & \multicolumn{3}{l}{Underflow concentration at the current step}                       \\
    \multirow[t]{5}{*}{Splits}
           & 5 training mild-jitter scenarios                                   & 1,355  & samples \\
           & 1 validation mild-jitter scenario                                  & 271    & samples \\
           & 1 near-domain test scenario                                        & 271    & samples \\
           & 12 far-domain test scenarios                                       & 3,252  & samples \\
           & 9 corrupted test cases                                             & 2,439  & samples \\
    \midrule
    \multicolumn{4}{l}{\textit{IndPenSim}}                                                         \\
    \multirow[t]{3}{*}{Input}
           & Variables                                                          & 23     &         \\
           & Sampling interval                                                  & 12     & minutes \\
           & Input window                                                       & 120    & steps   \\
    Target & \multicolumn{3}{l}{Assay-time penicillin concentration}                               \\
    \multirow[t]{6}{*}{Splits}
           & 24 training recipe-controlled batches                              & 451    & samples \\
           & 5 validation recipe-controlled batches                             & 90     & samples \\
           & 1 same-family recipe-controlled test batch                         & 19     & samples \\
           & 2 near-domain operator-controlled test batches                     & 38     & samples \\
           & 12 far-domain advanced process control (APC) or fault test batches & 220    & samples \\
           & 43 corrupted test cases                                            & 817    & samples \\
    \bottomrule
  \end{tabular}
\end{table}

\begin{figure*}[t]
  \centering
  \includegraphics[width=173mm]{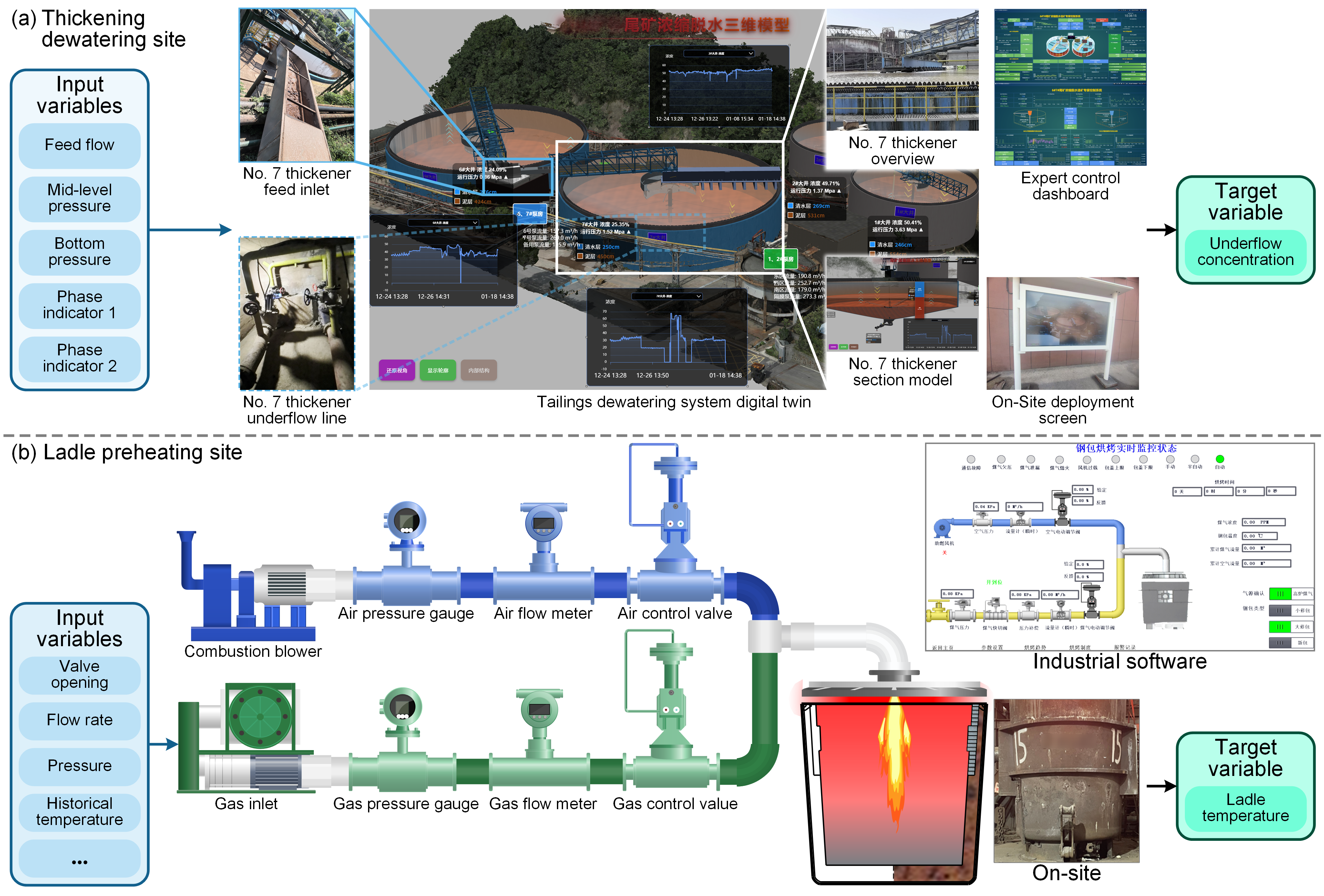}
  \caption{Industrial processes used in the experiments. (a)~Tailings thickening dewatering site in Nanjing, China. (b)~Ladle preheating site at a steel plant in Maanshan, China.}
  \label{fig:industrial_sites}
\end{figure*}

The ladle preheating dataset comes from a steel-metallurgy plant process that has also been used in variable ladle-preheating trend prediction~\cite{zong2025metacontrastive}. The task predicts future ladle inner temperature from fuel-gas, air-side, combustion-state, and historical-temperature variables. The main difficulty is complete-process holdout: each test process is an independent preheating run, so the model cannot rely on neighboring windows from the same process. The dataset has no curated regime or variable descriptions, so it also tests whether MCC can form usable measurement semantics under weak documentation.
The thickener dewatering dataset comes from a tailings thickening dewatering process. The task estimates current underflow concentration from feed, pressure, and operating-phase variables. Training data come from mild-jitter scenarios, while Real Test contains both a same-family near-domain scenario and far-domain scenarios. The far-domain scenarios cover visible measurement changes and hidden process changes, such as sensor bias, signal delay, load changes, slurry-property changes, and actuator-response changes.
The IndPenSim dataset is drawn from the public IndPenSim V3 penicillin fed-batch fermentation simulator~\cite{goldrick2015development}. The task estimates offline penicillin concentration from the online process window before each assay time. Training batches use recipe-controlled operation, while Real Test extends to same-family recipe, operator control, APC control, and fault batches. Thus, this dataset mainly tests soft-sensing stability under control-policy changes and fault conditions.

Corrupted Test is constructed separately from Real Test. It corrupts only the test subsets listed in \cref{tab:dataset_protocol} and does not change the training set, validation set, observer fitting, calibration, or predictor training. Single-variable corruptions are applied to selected correctable input variables, while global shift offsets all correctable input variables in the window. \cref{tab:corruption_protocol} defines the corruption types.

\begin{table}[t]
  \centering
  \caption{Test-time measurement-corruption types for Corrupted Test.}
  \label{tab:corruption_protocol}
  \setlength{\tabcolsep}{2pt}\tiny
  \begin{tabular}{ll}
    \toprule
    Corruption type & Measurement meaning \\
    \midrule
    Bias         & Constant offset in a selected measurement channel \\
    Drift        & Gradual calibration drift within the input window \\
    Gain         & Multiplicative scaling error in a selected measurement channel \\
    Spike        & Transient impulse at the current step \\
    Dropout      & Missing or zeroed measurement values \\
    Global shift & Simultaneous offset of all correctable input variables in the window \\
    \bottomrule
  \end{tabular}
\end{table}

\subsection{Main Results}\label{sec:4-3}

\cref{tab:main_ladle}, \cref{tab:main_thickener}, and \cref{tab:main_indpensim} report Base and +MCC on the three datasets. Real Test corresponds to the fixed real-test split, while Corrupted Test corresponds to the test subsets with controlled measurement corruption.

\begin{table}[t]
  \centering
  \caption{Quantitative comparison on the ladle preheating dataset. Base and +MCC are evaluated under the same Real Test and Corrupted Test protocols.}
  \label{tab:main_ladle}
  \setlength{\tabcolsep}{4pt}\tiny
  \begin{tabular}{llcccccc}
    \toprule
    \multirow{2}{*}{Backbone} & \multirow{2}{*}{Setting} & \multicolumn{3}{c}{Real Test} & \multicolumn{3}{c}{Corrupted Test} \\
    \cmidrule(lr){3-5}\cmidrule(lr){6-8}
                              &                          & MAE$\downarrow$ & RMSE$\downarrow$ & $R^2$$\uparrow$ & MAE$\downarrow$ & RMSE$\downarrow$ & $R^2$$\uparrow$ \\
    \midrule
    \multirow{2}{*}{GRU} & Base & 0.0209 & 0.0311 & 0.990 & 0.2285 & 0.2358 & 0.087 \\
                              & +MCC & 0.0158 & 0.0307 & 0.991 & 0.0175 & 0.0293 & 0.992 \\
    \multirow{2}{*}{LSTM} & Base & 0.0196 & 0.0324 & 0.990 & 0.1912 & 0.2028 & 0.283 \\
                              & +MCC & 0.0110 & 0.0208 & 0.996 & 0.0142 & 0.0245 & 0.994 \\
    \multirow{2}{*}{Transformer} & Base & 0.0269 & 0.0431 & 0.983 & 0.1828 & 0.1971 & 0.398 \\
                              & +MCC & 0.0155 & 0.0293 & 0.992 & 0.0186 & 0.0303 & 0.991 \\
    \multirow{2}{*}{Informer} & Base & 0.0290 & 0.0458 & 0.979 & 0.2902 & 0.3508 & -0.333 \\
                              & +MCC & 0.0154 & 0.0235 & 0.994 & 0.0191 & 0.0289 & 0.992 \\
    \multirow{2}{*}{Mamba} & Base & 0.0235 & 0.0374 & 0.986 & 0.1693 & 0.1957 & 0.492 \\
                              & +MCC & 0.0148 & 0.0252 & 0.994 & 0.0184 & 0.0293 & 0.992 \\
    \multirow{2}{*}{iTransformer} & Base & 0.0229 & 0.0385 & 0.986 & 0.1885 & 0.2079 & 0.338 \\
                              & +MCC & 0.0115 & 0.0180 & 0.997 & 0.0137 & 0.0208 & 0.996 \\
    \multirow{2}{*}{PatchTST} & Base & 0.0519 & 0.0878 & 0.924 & 0.1535 & 0.1845 & 0.530 \\
                              & +MCC & 0.0227 & 0.0361 & 0.987 & 0.0243 & 0.0385 & 0.985 \\
    \multirow{2}{*}{ModernTCN} & Base & 0.0270 & 0.0427 & 0.983 & 0.1724 & 0.1895 & 0.508 \\
                              & +MCC & 0.0240 & 0.0379 & 0.987 & 0.0259 & 0.0379 & 0.986 \\
    \bottomrule
  \end{tabular}
\end{table}

\begin{table}[t]
  \centering
  \caption{Quantitative comparison on the thickener dewatering dataset. Base and +MCC are evaluated under the same Real Test and Corrupted Test protocols.}
  \label{tab:main_thickener}
  \setlength{\tabcolsep}{4pt}\tiny
  \begin{tabular}{llcccccc}
    \toprule
    \multirow{2}{*}{Backbone} & \multirow{2}{*}{Setting} & \multicolumn{3}{c}{Real Test} & \multicolumn{3}{c}{Corrupted Test} \\
    \cmidrule(lr){3-5}\cmidrule(lr){6-8}
                              &                          & MAE$\downarrow$ & RMSE$\downarrow$ & $R^2$$\uparrow$ & MAE$\downarrow$ & RMSE$\downarrow$ & $R^2$$\uparrow$ \\
    \midrule
    \multirow{2}{*}{GRU} & Base & 0.0147 & 0.0312 & 0.211 & 0.0080 & 0.0081 & 0.621 \\
                              & +MCC & 0.0144 & 0.0314 & 0.201 & 0.0017 & 0.0020 & 0.943 \\
    \multirow{2}{*}{LSTM} & Base & 0.0148 & 0.0312 & 0.213 & 0.0086 & 0.0087 & 0.571 \\
                              & +MCC & 0.0145 & 0.0314 & 0.202 & 0.0004 & 0.0005 & 0.998 \\
    \multirow{2}{*}{Transformer} & Base & 0.0159 & 0.0315 & 0.198 & 0.0100 & 0.0104 & 0.363 \\
                              & +MCC & 0.0144 & 0.0301 & 0.209 & 0.0022 & 0.0027 & 0.933 \\
    \multirow{2}{*}{Informer} & Base & 0.0147 & 0.0315 & 0.195 & 0.0082 & 0.0095 & 0.426 \\
                              & +MCC & 0.0136 & 0.0304 & 0.201 & 0.0004 & 0.0005 & 0.999 \\
    \multirow{2}{*}{Mamba} & Base & 0.0161 & 0.0318 & 0.184 & 0.0073 & 0.0077 & 0.652 \\
                              & +MCC & 0.0153 & 0.0316 & 0.190 & 0.0008 & 0.0010 & 0.991 \\
    \multirow{2}{*}{iTransformer} & Base & 0.0156 & 0.0318 & 0.181 & 0.0048 & 0.0062 & 0.772 \\
                              & +MCC & 0.0138 & 0.0305 & 0.195 & 0.0011 & 0.0014 & 0.977 \\
    \multirow{2}{*}{PatchTST} & Base & 0.0155 & 0.0317 & 0.188 & 0.0068 & 0.0083 & 0.589 \\
                              & +MCC & 0.0140 & 0.0304 & 0.198 & 0.0004 & 0.0006 & 0.997 \\
    \multirow{2}{*}{ModernTCN} & Base & 0.0159 & 0.0316 & 0.194 & 0.0060 & 0.0065 & 0.756 \\
                              & +MCC & 0.0125 & 0.0292 & 0.238 & 0.0012 & 0.0014 & 0.977 \\
    \bottomrule
  \end{tabular}
\end{table}

\begin{table}[t]
  \centering
  \caption{Quantitative comparison on the IndPenSim dataset. Base and +MCC are evaluated under the same Real Test and Corrupted Test protocols.}
  \label{tab:main_indpensim}
  \setlength{\tabcolsep}{4pt}\tiny
  \begin{tabular}{llcccccc}
    \toprule
    \multirow{2}{*}{Backbone} & \multirow{2}{*}{Setting} & \multicolumn{3}{c}{Real Test} & \multicolumn{3}{c}{Corrupted Test} \\
    \cmidrule(lr){3-5}\cmidrule(lr){6-8}
                              &                          & MAE$\downarrow$ & RMSE$\downarrow$ & $R^2$$\uparrow$ & MAE$\downarrow$ & RMSE$\downarrow$ & $R^2$$\uparrow$ \\
    \midrule
    \multirow{2}{*}{GRU} & Base & 2.714 & 4.128 & 0.779 & 3.020 & 3.549 & 0.448 \\
                              & +MCC & 1.716 & 2.295 & 0.941 & 0.823 & 1.123 & 0.948 \\
    \multirow{2}{*}{LSTM} & Base & 2.327 & 3.437 & 0.842 & 2.496 & 3.169 & 0.530 \\
                              & +MCC & 1.283 & 1.783 & 0.929 & 0.925 & 1.373 & 0.921 \\
    \multirow{2}{*}{Transformer} & Base & 2.566 & 3.726 & 0.698 & 2.937 & 3.963 & 0.292 \\
                              & +MCC & 1.184 & 1.603 & 0.954 & 0.658 & 0.878 & 0.968 \\
    \multirow{2}{*}{Informer} & Base & 4.021 & 4.906 & 0.363 & 7.363 & 7.940 & -2.512 \\
                              & +MCC & 2.045 & 2.471 & 0.823 & 1.635 & 1.963 & 0.833 \\
    \multirow{2}{*}{Mamba} & Base & 3.118 & 4.623 & 0.742 & 3.548 & 4.445 & 0.055 \\
                              & +MCC & 1.353 & 1.813 & 0.923 & 1.156 & 1.506 & 0.906 \\
    \multirow{2}{*}{iTransformer} & Base & 4.781 & 6.309 & -0.087 & 5.317 & 6.629 & -0.804 \\
                              & +MCC & 3.634 & 5.358 & 0.647 & 3.826 & 4.688 & 0.100 \\
    \multirow{2}{*}{PatchTST} & Base & 2.708 & 3.971 & 0.795 & 2.758 & 3.458 & 0.471 \\
                              & +MCC & 1.801 & 2.473 & 0.751 & 1.067 & 1.416 & 0.913 \\
    \multirow{2}{*}{ModernTCN} & Base & 2.714 & 3.941 & 0.811 & 3.278 & 3.802 & 0.324 \\
                              & +MCC & 1.173 & 1.584 & 0.938 & 0.927 & 1.225 & 0.938 \\
    \bottomrule
  \end{tabular}
\end{table}

Across the 24 dataset--backbone pairs, +MCC reduces MAE relative to Base on both Real Test and Corrupted Test. Averaged over individual pairs, the relative MAE reduction is 30.7\% on Real Test and 80.3\% on Corrupted Test. Because Real Test contains no injected corruption, the gain is not restricted to synthetic recovery; MCC also remains useful on fixed industrial test splits.
At the dataset level, ladle preheating, thickener dewatering, and IndPenSim obtain average relative MAE reductions of 89.9\%, 85.9\%, and 64.9\% on Corrupted Test, respectively. The larger corrupted-test gains match the role of MCC: pre-inference correction is most valuable when local measurement distortion enters the input window. The smaller but consistent Real Test gains point to a conservative correction step rather than a tradeoff that sacrifices real-test performance for corruption recovery.

\subsection{Qualitative Analysis}\label{sec:4-4}

\begin{figure*}[t]
  \centering
  \includegraphics[width=\textwidth]{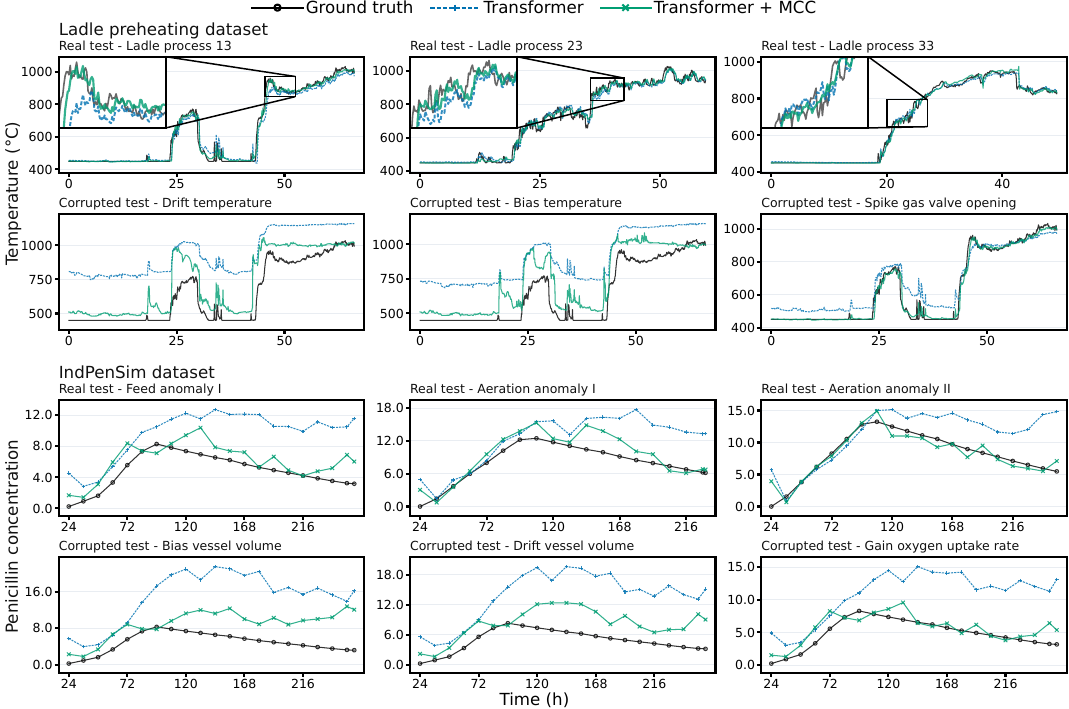}
  \caption{Representative Transformer prediction trajectories. The panels compare Base, +MCC, and target curves under the same Real Test and Corrupted Test protocols as the main tables.}
  \label{fig:qualitative_results}
\end{figure*}

\cref{fig:qualitative_results} shows representative prediction trajectories of Transformer and Transformer+MCC on Real Test and Corrupted Test.

On Real Test, the main effect of MCC is to reduce offsets rather than reshape the target trajectory. In ladle processes 13 and 23, MCC reduces deviations near stage transitions and local peaks; in process 33, the two prediction curves remain close, which suggests that MCC does not introduce a clear extra disturbance when no strong local conflict is present. In the three IndPenSim fault batches, Transformer tends to keep a persistent overestimation after the peak, while MCC pulls the trajectory back toward the decreasing trend and reduces the soft-sensing offset under fault operation.
On Corrupted Test, the difference follows the injected channel errors. Temperature drift and temperature bias shift the ladle baseline prediction upward, and MCC reduces this erroneous offset; the spike in gas valve opening is suppressed, so the prediction returns closer to the true temperature trajectory. In IndPenSim, vessel-volume bias, vessel-volume drift, and oxygen-uptake-rate gain cause the baseline to overestimate penicillin concentration over a long horizon, whereas MCC keeps the trajectory closer to the ground truth within the batch. This pattern is consistent with the Corrupted Test gains in the main results and points to the main role of MCC: weakening the propagation of local measurement distortion to the output trajectory.

\subsection{Ablation Studies}\label{sec:4-5}

\cref{tab:ablation} compares full MCC with key variants that weaken measurement semantics, variable--semantic match, reference routing, support and reliability, shared-shift suppression, and the correction rule.

\begin{table*}[t]
  \centering
  \caption{Ablation study on MCC design choices. Each row removes or replaces one component; MAE is reported for ModernTCN and Transformer, with lower values preferred.}
  \label{tab:ablation}
  {\setlength{\tabcolsep}{4pt}\tiny
    \begin{tabular}{@{}cccccccccccccc@{}}
      \toprule
      \multicolumn{6}{c}{Ablation setting} & \multicolumn{4}{c}{IndPenSim} & \multicolumn{4}{c}{Thickener Dewatering} \\
      \cmidrule(l{0pt}r{2pt}){1-6}\cmidrule(l{2pt}r{2pt}){7-10}\cmidrule(l{2pt}r{0pt}){11-14}
      \multirow{2}{*}[-0.8ex]{\makecell{Semantics}} & \multirow{2}{*}[-0.8ex]{\makecell{Semantic\\match}} & \multirow{2}{*}[-0.8ex]{\makecell{Reference}} & \multirow{2}{*}[-0.8ex]{\makecell{Reliable\\support}} & \multirow{2}{*}[-0.8ex]{\makecell{Shift\\suppression}} & \multirow{2}{*}[-0.8ex]{\makecell{Correction\\mode}} & \multicolumn{2}{c}{ModernTCN} & \multicolumn{2}{c}{Transformer} & \multicolumn{2}{c}{ModernTCN} & \multicolumn{2}{c}{Transformer} \\
      \cmidrule(lr){7-8}\cmidrule(lr){9-10}\cmidrule(lr){11-12}\cmidrule(lr){13-14}
       & & & & & & Real & Corrupted & Real & Corrupted & Real & Corrupted & Real & Corrupted \\
      \midrule
      --        & --         & --          & --         & --         & Raw input  & 2.7144 & 3.2785 & 2.5669 & 2.9372 & 0.0159 & 0.0060 & 0.0159 & 0.0100 \\
      LLM cards & Shuffled   & Semantic    & \checkmark & \checkmark & Calibrated & 2.3263 & 2.6363 & 1.4393 & 1.5591 & 0.0146 & 0.0020 & 0.0152 & 0.0031 \\
      Metadata  & \checkmark & Semantic    & \checkmark & \checkmark & Calibrated & 2.2234 & 2.0917 & 1.2743 & 0.9902 & 0.0139 & 0.0015 & 0.0150 & 0.0027 \\
      LLM cards & \checkmark & Correlation & \checkmark & \checkmark & Calibrated & 1.7589 & 1.8221 & 1.7640 & 1.2721 & 0.0135 & 0.0019 & 0.0150 & 0.0035 \\
      LLM cards & \checkmark & Semantic    & \checkmark & \checkmark & Residual   & 1.6750 & 1.9202 & 1.3027 & 1.3346 & 0.0137 & 0.0016 & 0.0151 & 0.0032 \\
      LLM cards & \checkmark & Semantic    & --         & \checkmark & Calibrated & 1.8181 & 1.3542 & 1.7661 & 1.6538 & 0.0133 & 0.0014 & 0.0151 & 0.0026 \\
      LLM cards & \checkmark & Semantic    & \checkmark & --         & Calibrated & 2.0405 & 1.3740 & 1.5222 & 1.1816 & 0.0142 & 0.0013 & 0.0153 & 0.0027 \\
      LLM cards & \checkmark & Semantic    & \checkmark & \checkmark & Direct     & 2.4432 & 1.8126 & 2.1188 & 1.9880 & 0.0129 & 0.0024 & 0.0148 & 0.0024 \\
      LLM cards & \checkmark & Semantic    & \checkmark & \checkmark & Calibrated & 1.1738 & 0.9276 & 1.1841 & 0.6583 & 0.0125 & 0.0012 & 0.0144 & 0.0022 \\
      \bottomrule
    \end{tabular}
  }
\end{table*}

Full MCC obtains the lowest MAE in all representative dataset--backbone--protocol combinations, so the main gain is not tied to a single backbone or a single test protocol. The advantage is clearest on IndPenSim. On the thickener Real Test, the baseline error is already small; full MCC still ranks first, but with a narrower margin.

The semantic ablations separate the role of measurement semantics from observer capacity. Shuffling the semantic match or replacing LLM cards with metadata weakens the result, and correlation-based references do not consistently replace semantic references. The conclusion is that external evidence should be qualified by measurement meaning rather than selected only by numerical association.

The correction-rule ablations show that the gain comes from constrained local correction rather than a looser repair strategy. Residual triggering, removing reliable support, removing shift suppression, and direct reference replacement all underperform the full design. Direct replacement is close to full MCC on the thickener Real Test, but it is unstable on corrupted tests and on IndPenSim. Thus, full MCC does not pull every deviation toward a reference. It applies calibrated and conservative local correction when reliable external evidence supports the conflict.

\subsection{MCC Correction Mechanism Visualization}\label{sec:4-6}

\begin{figure*}[t]
  \centering
  \includegraphics[width=\textwidth]{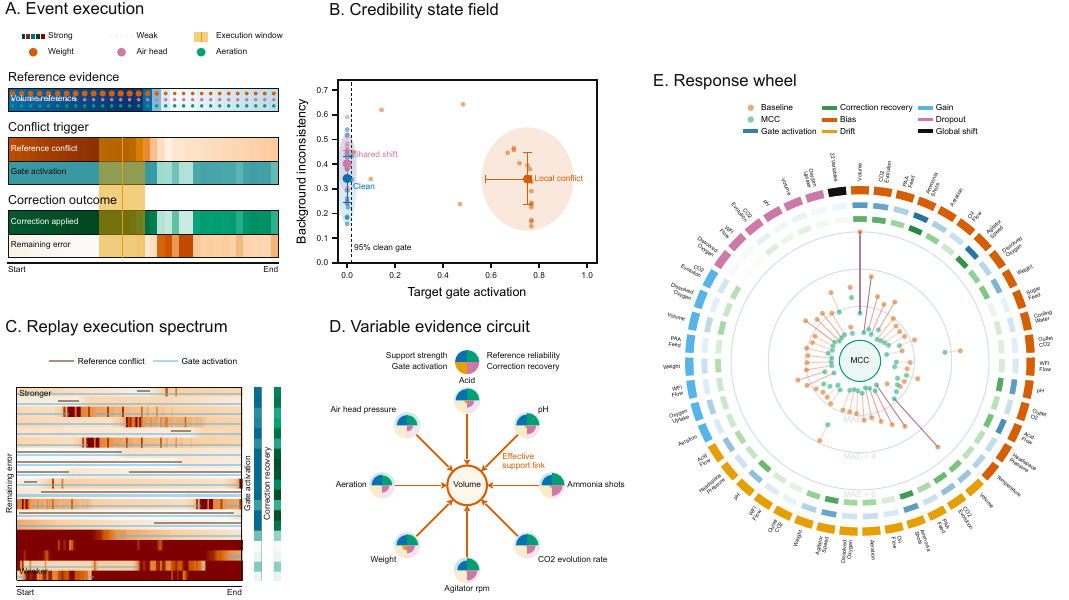}
  \caption{MCC correction mechanism on IndPenSim with Transformer. (a) Event replay for a local \texttt{volume} bias. (b) Gate activation versus background inconsistency for clean, local-conflict, and shared-shift windows. (c) Replay spectrum for local \texttt{volume} corruptions, with remaining error in the heatmap and gate activation/recovery on the side rails. (d) Evidence circuit for \texttt{volume}. (e) Controlled-corruption response wheel comparing Base and MCC MAE.}
  \label{fig:mcc_correction_mechanism_visualization}
\end{figure*}

\cref{fig:mcc_correction_mechanism_visualization} examines the internal correction mechanism of MCC on the IndPenSim Transformer run, focusing on qualified external evidence, local measurement conflicts, and conservative behavior under shared operating movement.

For \texttt{volume}, the main supporting entries come from \texttt{weight}, \texttt{air\_head\_pressure}, and \texttt{aeration}, rather than from the target measurement itself. Under local \texttt{volume} bias, the measurement error decreases from 1.500 to 0.545, and the prediction MAE decreases from 6.322 for the baseline to 1.634 for MCC. The event is driven by an externally supported reference, not by mean regression or unconstrained reconstruction.

The figure also gives the boundary of selective correction. For \texttt{volume}, the mean gate activation is only 0.002 on clean windows, and both the mean gate activation and correction gain are 0.000 under the shared 22-variable shift; shared operating movement is not treated as a local \texttt{volume} distortion. Across 47 non-clean controlled corruption cases, the average MAE decreases from 2.363 to 1.047, and MCC outperforms the baseline in 46 cases; the only non-improved case is the local \texttt{temp} bias.

\subsection{Sensitivity Analysis}\label{sec:4-7}

\begin{figure*}[t]
  \centering
  \includegraphics[width=\linewidth]{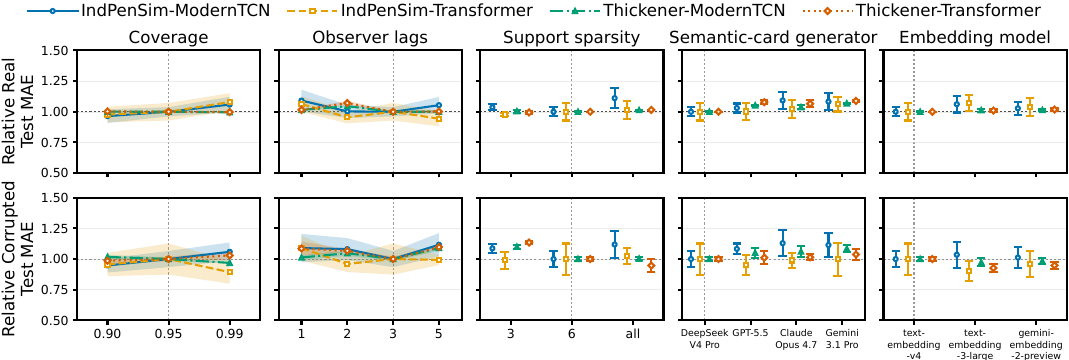}
  \caption{Sensitivity of MCC to main design choices. Values are normalized by the default setting; horizontal and vertical dashed lines mark default performance and configurations.}
  \label{fig:sensitivity_analysis}
\end{figure*}

\cref{fig:sensitivity_analysis} evaluates the sensitivity of MCC to key design choices, including calibration coverage, observer lag order, support sparsity, semantic-card generator, and embedding model.

The sensitivity results do not show a narrow single-point operating region. Calibration coverage, observer lag order, and support sparsity remain stable around the default point; after changing the LLM generator or the embedding model, the relative MAE also stays mostly close to 1.0. The main gain of MCC appears to come from measurement-semantic structure rather than one specific model output or a finely tuned single setting.

\subsection{Computational Efficiency}\label{sec:4-8}

To evaluate deployment cost, \cref{tab:efficiency} separates offline semantic construction from online inference and reports model size, per-step inference time, and peak-memory change after adding MCC to the same backbone.

\begin{table}[t]
  \centering
  \caption{Online deployment cost of MCC. The table counts only online parameters, inference time, and Peak RSS; offline LLM semantic generation and embedding construction are excluded from per-step inference.}
  \label{tab:efficiency}
  \setlength{\tabcolsep}{4pt}\tiny
  \begin{tabular}{lrrrrr}
    \toprule
              & \multicolumn{2}{c}{Model Size} & \multicolumn{3}{c}{Deployment Cost} \\
    \cmidrule(lr){2-3}\cmidrule(lr){4-6}
    Backbone  & \makecell{Backbone\\(\,\texttimes\,10\textsuperscript{3})} & \makecell{MCC Inc.\\(\,\texttimes\,10\textsuperscript{3})} & \makecell{Base Inference\\Time (ms/step)} & \makecell{+MCC Inference\\Time (ms/step)} & \makecell{Peak RSS\\Increase (MB)} \\
    \midrule
    \multicolumn{6}{l}{\textit{IndPenSim}} \\
    ModernTCN   & 78.8  & 2.0 & 0.015 & 0.088 & 284.8 \\
    Transformer & 188.4 & 2.0 & 0.015 & 0.089 & 299.2 \\
    \addlinespace[1pt]
    \multicolumn{6}{l}{\textit{Thickener}} \\
    ModernTCN   & 77.1  & 0.5 & 0.013 & 0.041 & 289.1 \\
    Transformer & 178.0 & 0.5 & 0.013 & 0.041 & 303.6 \\
    \bottomrule
  \end{tabular}
\end{table}

The table reports the online part of MCC; LLM semantic generation and embedding construction are completed before deployment. Online, the observer adds only 0.5--2.0k parameters; the slowest +MCC inference time is 0.089 ms/step, the largest added latency is 0.074 ms/step, and the Peak RSS increase stays below 304 MB. These costs are small relative to the 1--12 minute sampling intervals in the datasets. Overall, MCC adds the credibility-correction step with a small online cost.

\section{Conclusion}\label{sec:conclusion}

This article presented MCC for pre-inference measurement credibility correction in industrial process inference. MCC uses LLM-derived measurement semantics to decide which external variables can provide credible evidence for a target measurement. It then corrects supported local measurement conflicts before prediction, so the downstream predictor receives a more credible input window.
Across multiple complex industrial forecasting and soft-sensing tasks, MCC reduces prediction error, with average relative MAE reductions of 30.7\% and 80.3\% on real-test and controlled-corruption protocols, respectively. MCC works with different backbones and keeps strong performance. Online, it adds only 0.5--2.0k parameters, with the slowest inference time at 0.089 ms/step, making it lightweight for deployment.
The main boundary of MCC is whether available process documents correctly reflect the field measurement chain. When variable descriptions are incomplete, outdated, or inconsistent, measurement semantics should be validated with field records. Future work will study long-term semantic validation and credibility-correction updates when measurement chains change.

\bibliographystyle{IEEEtran}
\bibliography{references}

@article{sun2021survey,
  author  = {Sun, Qingqiang and Ge, Zhiqiang},
  journal = {IEEE Transactions on Industrial Informatics},
  title   = {A Survey on Deep Learning for Data-Driven Soft Sensors},
  year    = {2021},
  volume  = {17},
  number  = {9},
  pages   = {5853--5866},
  doi     = {10.1109/TII.2021.3053128}
}

@article{yuan2020lstm,
  author  = {Yuan, Xiaofeng and Li, Lin and Wang, Yalin},
  journal = {IEEE Transactions on Industrial Informatics},
  title   = {Nonlinear Dynamic Soft Sensor Modeling With Supervised Long Short-Term Memory Network},
  year    = {2020},
  volume  = {16},
  number  = {5},
  pages   = {3168--3176},
  doi     = {10.1109/TII.2019.2902129}
}

@article{geng2022transformer,
  author  = {Geng, Zhiqiang and Chen, Zhiwei and Meng, Qingchao and Han, Yongming},
  journal = {IEEE Transactions on Industrial Informatics},
  title   = {Novel Transformer Based on Gated Convolutional Neural Network for Dynamic Soft Sensor Modeling of Industrial Processes},
  year    = {2022},
  volume  = {18},
  number  = {3},
  pages   = {1521--1529},
  doi     = {10.1109/TII.2021.3086798}
}

@inproceedings{vaswani2017transformer,
  author    = {Vaswani, Ashish and Shazeer, Noam and Parmar, Niki and Uszkoreit, Jakob and Jones, Llion and Gomez, Aidan N. and Kaiser, {\L}ukasz and Polosukhin, Illia},
  booktitle = {Advances in Neural Information Processing Systems 30 (NeurIPS 2017)},
  title     = {Attention Is All You Need},
  year      = {2017},
  pages     = {5998--6008},
  url       = {https://papers.neurips.cc/paper/7181-attention-is-all-you-need}
}

@inproceedings{liu2024itransformer,
  author    = {Liu, Yong and Hu, Tengge and Zhang, Haoran and Wu, Haixu and Wang, Shiyu and Ma, Lintao and Long, Mingsheng},
  booktitle = {Proceedings of the International Conference on Learning Representations (ICLR)},
  title     = {{iTransformer}: Inverted Transformers Are Effective for Time Series Forecasting},
  year      = {2024},
  doi       = {10.48550/arXiv.2310.06625},
  url       = {https://openreview.net/forum?id=JePfAI8fah}
}

@inproceedings{nie2023patchtst,
  author    = {Nie, Yuqi and Nguyen, Nam H. and Sinthong, Phanwadee and Kalagnanam, Jayant},
  booktitle = {Proceedings of the International Conference on Learning Representations (ICLR)},
  title     = {A Time Series Is Worth 64 Words: Long-Term Forecasting With Transformers},
  year      = {2023},
  doi       = {10.48550/arXiv.2211.14730},
  url       = {https://openreview.net/forum?id=Jbdc0vTOcol}
}

@inproceedings{luo2024moderntcn,
  author    = {Luo, Donghao and Wang, Xue},
  booktitle = {Proceedings of the International Conference on Learning Representations (ICLR)},
  title     = {{ModernTCN}: A Modern Pure Convolution Structure for General Time Series Analysis},
  year      = {2024},
  url       = {https://openreview.net/forum?id=vpJMJerXHU}
}

@inproceedings{gu2024mamba,
  author    = {Gu, Albert and Dao, Tri},
  booktitle = {The First Conference on Language Modeling (COLM)},
  title     = {Mamba: Linear-Time Sequence Modeling with Selective State Spaces},
  year      = {2024},
  url       = {https://openreview.net/forum?id=tEYskw1VY2}
}

@article{zong2026llmdecision,
  author  = {Zong, Youcheng and Jia, Runda and Li, Kang and Xue, Dazhan and Zhang, Liqiang and He, Dakuo},
  journal = {Neural Networks},
  title   = {{LLM}-Driven Human-{AI} Collaborative Decision Support System for Complex Industrial Processes: A Case Study in Metallurgy},
  year    = {2026},
  volume  = {202},
  pages   = {109055},
  doi     = {10.1016/j.neunet.2026.109055},
  url     = {https://www.sciencedirect.com/science/article/pii/S0893608026005150}
}

@article{liu2024kgllm,
  author  = {Liu, Peifeng and Qian, Lu and Zhao, Xingwei and Tao, Bo},
  journal = {IEEE Transactions on Industrial Informatics},
  title   = {Joint Knowledge Graph and Large Language Model for Fault Diagnosis and Its Application in Aviation Assembly},
  year    = {2024},
  volume  = {20},
  number  = {6},
  pages   = {8160--8169},
  doi     = {10.1109/TII.2024.3366977}
}

@inproceedings{merrill2024llmts,
  author    = {Tan, Mingtian and Merrill, Mike A. and Gupta, Vinayak and Althoff, Tim and Hartvigsen, Thomas},
  booktitle = {Advances in Neural Information Processing Systems 37},
  title     = {Are Language Models Actually Useful for Time Series Forecasting?},
  year      = {2024},
  volume    = {37},
  pages     = {60162--60191},
  doi       = {10.52202/079017-1922},
  url       = {https://proceedings.neurips.cc/paper_files/paper/2024/hash/6ed5bf446f59e2c6646d23058c86424b-Abstract-Conference.html}
}

@article{zong2025hybridgrid,
  author  = {Zong, Youcheng and Nian, Yi and Zhang, Chaojie and Tang, Xinyu and Wang, Lin and Zhang, LiQiang},
  title   = {Hybrid Grid Search and {Bayesian} Optimization-Based Random Forest Regression for Predicting Material Compression Pressure in Manufacturing Processes},
  journal = {Engineering Applications of Artificial Intelligence},
  year    = {2025},
  volume  = {141},
  pages   = {109580},
  doi     = {10.1016/j.engappai.2024.109580},
  url     = {https://www.sciencedirect.com/science/article/pii/S095219762401738X},
  issn    = {0952-1976}
}

@article{zong2025metacontrastive,
  author  = {Zong, Youcheng and Jia, Runda and Wu, Shuai and Zhang, Liqiang and He, Dakuo},
  title   = {A meta-contrastive learning hybrid model for adaptive temperature trend prediction in variable ladle preheating},
  journal = {Engineering Applications of Artificial Intelligence},
  year    = {2025},
  volume  = {162},
  pages   = {112750},
  issn    = {0952-1976},
  doi     = {10.1016/j.engappai.2025.112750},
  url     = {https://www.sciencedirect.com/science/article/pii/S0952197625027812}
}

@article{su2025zero,
  author  = {Su, Weilan and Zong, Youcheng and Jia, Runda and Qin, Jian and Li, Ming},
  journal = {IEEE Journal of Biomedical and Health Informatics},
  title   = {Zero-Shot Capillary Segmentation in Dermoscopy Images via {SAM2}: A Case Study on Oral Mucosa},
  year    = {2026},
  volume  = {30},
  number  = {5},
  pages   = {4376--4387},
  doi     = {10.1109/JBHI.2025.3628493},
  issn    = {2168-2208}
}

@article{yao2026causalllm,
  author  = {Yao, Fang and Liu, Jizhe and Tao, Yuechuan and Qiu, Jing and Iu, Herbert Ho-Ching and Chen, Guo and Dong, Zhao Yang},
  journal = {IEEE Transactions on Industrial Informatics},
  title   = {Causality-Aware {LLM}-Enhanced Graph Representation Learning for Adaptive Power System Control},
  year    = {2026},
  volume  = {22},
  number  = {5},
  pages   = {3681--3692},
  doi     = {10.1109/TII.2026.3651250}
}

@article{han2026zeroshot,
  author  = {Han, Huazheng and Gao, Xuejin and Han, Huayun and Gao, Huihui and Qi, Yongsheng},
  journal = {IEEE Transactions on Industrial Informatics},
  title   = {Zero-Shot Fault Diagnosis via {LLM}-Guided Complexity-Aware Fuzzy Boundary Learning},
  year    = {2026},
  pages   = {1--12},
  doi     = {10.1109/TII.2026.3669353}
}

@article{goldrick2015development,
  title={The development of an industrial-scale fed-batch fermentation simulation},
  author={Goldrick, Stephen and {\c{S}}tefan, Andrei and Lovett, David and Montague, Gary and Lennox, Barry},
  journal={Journal of Biotechnology},
  volume={193},
  pages={70--82},
  year={2015},
  publisher={Elsevier},
  doi={10.1016/j.jbiotec.2014.10.029}
}

@inproceedings{liu2025timecma,
  author    = {Liu, Chenxi and Xu, Qianxiong and Miao, Hao and Yang, Sun and Zhang, Lingzheng and Long, Cheng and Li, Ziyue and Zhao, Rui},
  booktitle = {Proceedings of the AAAI Conference on Artificial Intelligence},
  title     = {{TimeCMA}: Towards {LLM}-Empowered Multivariate Time Series Forecasting via Cross-Modality Alignment},
  year      = {2025},
  pages     = {18780--18788},
  doi       = {10.1609/AAAI.V39I18.34067},
  url       = {https://mlanthology.org/aaai/2025/liu2025aaai-timecma/}
}

@inproceedings{jin2024timellm,
  author    = {Jin, Ming and Wang, Shiyu and Ma, Lintao and Chu, Zhixuan and Zhang, James Y. and Shi, Xiaoming and Chen, Pin-Yu and Liang, Yuxuan and Li, Yuan-Fang and Pan, Shirui and Wen, Qingsong},
  booktitle = {Proceedings of the International Conference on Learning Representations (ICLR)},
  title     = {{Time-LLM}: Time Series Forecasting by Reprogramming Large Language Models},
  year      = {2024},
  doi       = {10.48550/arXiv.2310.01728},
  url       = {https://arxiv.org/abs/2310.01728}
}

@inproceedings{sun2024test,
  author    = {Sun, Chenxi and Li, Hongyan and Li, Yaliang and Hong, Shenda},
  booktitle = {Proceedings of the International Conference on Learning Representations (ICLR)},
  title     = {{TEST}: Text Prototype Aligned Embedding to Activate {LLM}'s Ability for Time Series},
  year      = {2024},
  doi       = {10.48550/arXiv.2308.08241},
  url       = {https://arxiv.org/abs/2308.08241}
}

@inproceedings{liu2024autotimes,
  author    = {Liu, Yong and Qin, Guo and Huang, Xiangdong and Wang, Jianmin and Long, Mingsheng},
  booktitle = {Advances in Neural Information Processing Systems 37},
  title     = {{AutoTimes}: Autoregressive Time Series Forecasters via Large Language Models},
  year      = {2024},
  doi       = {10.48550/arXiv.2402.02370},
  url       = {https://arxiv.org/abs/2402.02370}
}

@article{venkatasubramanian2003fddquantitative,
  author  = {Venkatasubramanian, Venkat and Rengaswamy, Raghunathan and Yin, Kewen and Kavuri, Surya N.},
  journal = {Computers \& Chemical Engineering},
  title   = {A Review of Process Fault Detection and Diagnosis},
  year    = {2003},
  volume  = {27},
  number  = {3},
  pages   = {293--311},
  doi     = {10.1016/S0098-1354(02)00160-6}
}

@article{qin2012processmonitoring,
  author  = {Qin, S. Joe},
  journal = {Annual Reviews in Control},
  title   = {Survey on Data-Driven Industrial Process Monitoring and Diagnosis},
  year    = {2012},
  volume  = {36},
  number  = {2},
  pages   = {220--234},
  doi     = {10.1016/j.arcontrol.2012.09.004}
}

@article{yue2001reconstruction,
  author  = {Yue, H. Henry and Qin, S. Joe},
  journal = {Industrial \& Engineering Chemistry Research},
  title   = {Reconstruction-Based Fault Identification Using a Combined Index},
  year    = {2001},
  volume  = {40},
  number  = {20},
  pages   = {4403--4414},
  doi     = {10.1021/ie000141+}
}

@book{narasimhan1999datareconciliation,
  author    = {Narasimhan, Shankar and Jordache, Cornelius},
  title     = {Data Reconciliation and Gross Error Detection: An Intelligent Use of Process Data},
  publisher = {Gulf Professional Publishing},
  year      = {1999},
  isbn      = {0884152553}
}

@article{zhai2024processgraphsoftsensor,
  author  = {Zhai, Ruikun and Zheng, Junhua and Song, Zhihuan and Ge, Zhiqiang},
  journal = {IEEE Transactions on Industrial Informatics},
  title   = {Reliable Soft Sensors With an Inherent Process Graph Constraint},
  year    = {2024},
  volume  = {20},
  number  = {6},
  pages   = {8798--8806},
  doi     = {10.1109/TII.2024.3372013}
}

@article{zhang2025faulttolerantsoftsensor,
  author  = {Zhang, Xiangrui and Song, Chunyue and Zhao, Jun and Huang, Biao},
  journal = {IEEE Transactions on Automation Science and Engineering},
  title   = {Fault-Tolerant Soft Sensor Modeling Based on a Two-Dimensional Group Distributionally Robust Optimization Framework},
  year    = {2025},
  volume  = {22},
  pages   = {14396--14406},
  doi     = {10.1109/TASE.2025.3561754}
}

@article{liu2026transfersoftsensor,
  author  = {Liu, Yi and Zhu, Jialiang and Yang, Chao and Chen, Tao and Wong, David Shan Hill and Yao, Yuan},
  journal = {Industrial \& Engineering Chemistry Research},
  title   = {Transfer Learning for Soft Sensors in Process Industries: A Review and Future Perspectives},
  year    = {2026},
  volume  = {65},
  number  = {16},
  pages   = {8103--8125},
  doi     = {10.1021/acs.iecr.5c05144}
}

@inproceedings{deng2021gnnad,
  author    = {Deng, Ailin and Hooi, Bryan},
  booktitle = {Proceedings of the AAAI Conference on Artificial Intelligence},
  title     = {Graph Neural Network-Based Anomaly Detection in Multivariate Time Series},
  year      = {2021},
  volume    = {35},
  number    = {5},
  pages     = {4027--4035},
  doi       = {10.1609/aaai.v35i5.16523}
}

@article{zheng2024adversarialgnn,
  author  = {Zheng, Bolong and Ming, Lingfeng and Zeng, Kai and Zhou, Mengtao and Zhang, Xinyong and Ye, Tao and Yang, Bin and Zhou, Xiaofang and Jensen, Christian S.},
  journal = {IEEE Transactions on Knowledge and Data Engineering},
  title   = {Adversarial Graph Neural Network for Multivariate Time Series Anomaly Detection},
  year    = {2024},
  volume  = {36},
  number  = {12},
  pages   = {7612--7626},
  doi     = {10.1109/TKDE.2024.3419891}
}

\end{document}